\documentclass[aps,prd,twocolumn,nofootinbib,floatfix,showpacs,showkeys,preprintnumbers,superscriptaddress]{revtex4}
\usepackage{subfigure}
\usepackage{graphicx}
\usepackage{dcolumn}
\usepackage{epsfig}
\usepackage{amsmath}
\usepackage{amsfonts}
\usepackage{amssymb}
\usepackage{color}
\usepackage{hyperref}
\usepackage{ulem}

\usepackage{hhline}
\usepackage{array}
\newcommand{\PreserveBackslash}[1]{\let\temp=\\#1\let\\=\temp}
\newcolumntype{C}[1]{>{\PreserveBackslash\centering}p{#1}}
\newcolumntype{R}[1]{>{\PreserveBackslash\raggedleft}p{#1}}
\newcolumntype{L}[1]{>{\PreserveBackslash\raggedright}p{#1}}

\newcommand{\LCDM}{$\Lambda$CDM}

\newcommand{\be}{\begin{equation}}
\newcommand{\ee}{\end{equation}}
\newcommand{\bea}{\begin{eqnarray}}
\newcommand{\eea}{\end{eqnarray}}
\newcommand{\nn}{\nonumber}
\newcommand{\di}{\text{d}}

\begin{document}

\title{Linear Point and Sound Horizon as Purely Geometric standard rulers}

\author{M\'arcio O'Dwyer}
\affiliation{Department of Physics/CERCA/Institute for the Science of Origins, Case Western Reserve University, Cleveland, OH 44106-7079 -- USA}

\author{Stefano Anselmi}
\affiliation{LUTH, UMR 8102 CNRS, Observatoire de Paris, PSL Research University, Universit\'e Paris Diderot, 92190 Meudon -- France}
\affiliation{Institut d'Astrophysique de Paris, CNRS UMR 7095 and UPMC, 98bis, bd Arago, F-75014 Paris -- France}
\affiliation{Department of Physics, Israel Institute of Technology Technion, Haifa 320003 -- Israel}
\email{stefano@campus.technion.ac.il}

\author{Glenn D.~Starkman}
\affiliation{Department of Physics/CERCA/Institute for the Science of Origins, Case Western Reserve University, Cleveland, OH 44106-7079 -- USA}

\author{Pier-Stefano Corasaniti}
\affiliation{LUTH, UMR 8102 CNRS, Observatoire de Paris, PSL Research University, Universit\'e Paris Diderot, 92190 Meudon -- France}
\affiliation{Institut d'Astrophysique de Paris, CNRS UMR 7095 and UPMC, 98bis, bd Arago, F-75014 Paris -- France}

\author{Ravi K.~Sheth}
\affiliation{Center for Particle Cosmology, University of Pennsylvania, 209 S. 33rd St., Philadelphia, PA 19104 -- USA}
\affiliation{The Abdus Salam International Center for Theoretical Physics, Strada Costiera, 11, Trieste 34151 -- Italy}

\author{Idit Zehavi}
\affiliation{Department of Physics/CERCA/Institute for the Science of Origins, Case Western Reserve University, Cleveland, OH 44106-7079 -- USA}

\date{\today}

\begin{abstract}
The Baryon Acoustic Oscillations feature (BAO) imprinted in the clustering correlation function is known to furnish us cosmic distance determinations that are independent of the cosmological-background model and the primordial perturbation parameters. These measurements can be accomplished rigorously by means of the Purely Geometric BAO methods. To date two different Purely Geometric BAO approaches have been proposed. The first exploits the {\it linear-point} standard ruler. The second, called correlation-function model-fitting, exploits the {\it sound-horizon} standard ruler. A key difference between them is that, when estimated from clustering data, the linear point makes use of a cosmological-model-independent procedure to extract the ratio of the ruler to the cosmic distance, while the correlation-function model-fitting relies on a phenomenological cosmological model for the correlation function. Nevertheless the two rulers need to be precisely defined independently of any specific observable (e.g.~the BAO). We define the linear point and sound horizon and we fully characterize and compare the two rulers' cosmological-parameter dependence. We find that they are both geometrical (i.e.~independent of the primordial cosmological parameters) within the required accuracy, and that they have the same parameter dependence for a wide range of parameter values. We  estimate the rulers' best-fit values and errors, given the cosmological constraints obtained by the Planck Satellite team from their measurements of the Cosmic Microwave Background temperature and polarization anisotropies. We do this for three different cosmological models encompassed by the Purely Geometric BAO methods. In each case we find that the relative errors of the two rulers coincide and they are insensitive to the assumed cosmological model. Interestingly both the linear point and the sound horizon shift by $0.5\sigma$ when we do not fix the spatial geometry to be flat in \LCDM. This points toward a sensitivity of the rulers to different cosmological models when they are estimated from the Cosmic Microwave Background. 
\end{abstract}


\maketitle


\section{Introduction}\label{sec:intro}
\label{intro}

In cosmology standard rulers are a valuable tool to infer cosmological information from observational data. In this respect the Baryon Acoustic Oscillations (BAO) are known to furnish us a powerful comoving cosmological standard ruler \cite{Bassett:2009mm}. In our standard cosmological description initial fluctuations in the gravitational potential drove acoustic waves in the primordial photon-baryon plasma. Soon after the decoupling time, at the drag epoch, the acoustic waves stopped propagating leaving in the baryon distribution  overdensities separated by a characteristic length scale -- the sound-horizon comoving length at the drag epoch $r_{d}$ (SH). The subsequent large-scale evolution of matter (i.e.~dark matter and baryons) is dominated by gravity that left in the 2-point correlation functions (CF) of the matter and its measured tracers' (e.g.~galaxies) a characteristic feature that we call Baryon Acoustic Oscillations \cite{Eisenstein:2005su}.

Since the CF BAO feature was initially identified with the so-called acoustic peak, and since its position is close to the sound-horizon scale, the original idea was to measure the same length-scale from the primordial and late Universe and exploit it for cosmology. However, in this era of precision cosmology, this excellent intuition encountered certain challenges. Firstly, we now know that generically we cannot estimate the primordial $r_{d}$ in a cosmology-model-independent way and without modeling the late-time physics\footnote{See, however, \cite{2017JCAP...04..023V} for a discussion of how  late-time effects can be removed from Cosmic Microwave Background data for cosmological models that are sufficiently close to flat-$\Lambda$CDM.}
. Secondly, to extract $r_{d}$ from late-time data, we need to  model the galaxy-correlation-function non-linearities that squash and shift the peak position in a time-dependent and model-dependent way \cite{2008PhRvD..77d3525S, 2008MNRAS.390.1470S}. 

Hence nowadays we adopt a specific (and somewhat arbitrary) definition for the sound horizon scale $r_{d}$ \cite{Thepsuriya:2014zda}. Assuming a cosmological model (e.g. flat \LCDM) to fit the data from the Cosmic Microwave Background (CMB) anisotropy power spectrum data, $r_{d}$ is estimated as a secondary (i.e.~derived) parameter (e.g.~\cite{2016A&A...594A..13P}). At late times $r_{d}$ again needs to be estimated as a secondary parameter derived from a cosmology-dependent multi-parameter fit  to the galaxy-correlation-function data \cite{2019PhRvD..99l3515A}. To be more precise, from BAO measurements, the interesting and best-measured quantity is the ratio of the sound-horizon scale to the cosmic distance of the galaxy survey \cite{2019PhRvD..99l3515A}. 

To use the BAO as a standard ruler, the BAO distance measures (in units of the SH scale) must be estimated with ``Purely Geometric-BAO (PG-BAO) methods,'' as described in \cite{2019PhRvD..99l3515A}. In these methods one assumes neither a flat spatial geometry nor a specific model for the late-time cosmic acceleration. The PG-BAO methods require the estimated distance measures to be geometrical, i.e.~independent of the primordial-fluctuation parameters. Such PG-BAO measurements are one of the main motivations behind the large effort devoted to build  upcoming Large Scale Structure (LSS) galaxy surveys such as Euclid,\footnote{\url{http://sci.esa.int/euclid/}} DESI,\footnote{\url{http://desi.lbl.gov}} and WFIRST\footnote{\url{https://wfirst.gsfc.nasa.gov}}.

The standard BAO approach to estimate distances \cite{2008ApJ...686...13S, 2012MNRAS.427.2146X}, called {\it BAO-Only}, was intended to be a PG-BAO method. However it fails to meet the requirements, as the cosmological parameters and the non-linear damping  parameters are kept fixed in the CF template fit \cite{2014MNRAS.441...24A}. Notice that even standard BAO state-of-the-art methods and tests do not address and solve this issue that remains a serious limitation of the standard BAO approach \cite{2019arXiv190603035C}.  Because of this parameter fixing, the sound-horizon scale is not estimated as a secondary parameter but instead enters as an {\it ad hoc} correction term necessary to obtain unbiased results from the CF fits validated with simulations or survey mocks (see Section 7 of \cite{2008ApJ...686...13S}). To overcome this uncontrolled approximation, recently some of us proposed a new PG-BAO approach named correlation-function model-fitting (CF-MF) \cite{2019PhRvD..99l3515A}. CF-MF allows to properly propagate all the uncertainties without fixing the cosmological parameters. The sound-horizon scale is properly obtained as a secondary parameter from the fit. Using CF-MF, it was possible to show that, when the BAO-Only technique is employed to infer the ratio of the sound-horizon scale to the cosmic distance, a Fisher-matrix analysis finds that it underestimates the errors by up to a factor of 2 unless one imposes CMB-related priors on the parameters \cite{2019PhRvD..99l3515A}. 

CF-MF relies on a CF template: a phenomenological model of the galaxy correlation function. In fact we are not able to unambiguously predict the galaxy CF starting from cosmological initial conditions. As a matter of fact, many CF templates have been proposed in the past and it is not clear which one to choose, how many parameters are needed to describe galaxy clustering, and on which range of scales the fit to the data should be performed \cite{2019PhRvD..99f3530O}. The lack of a fundamental model of the predicted CF forces one to tune the CF-templates to the simulated outcomes. Since the N-body simulations are always run for cosmological parameter values close to the best-fit Planck posteriors, the chosen template, when employed for galaxy-data fits, could bias the best-fit inferred distances and/or errors. 

To overcome the theoretical limitations of the CF-MF approach, recently a complementary new point of view to perform PG-BAO measurements was proposed. A new standard ruler in the BAO range of scales was identified and dubbed the {\it Linear Point} (LP). It is defined as the mid-point between the peak and the dip positions of the CF computed in linear theory in the BAO range of scales. It was proven \cite{2016MNRAS.455.2474A} to be geometrical near the Planck-2015 best-fit flat-\LCDM~cosmology \cite{2016A&A...594A..13P}. However, what makes the LP really valuable is its insensitivity (at the $0.5\%$ level) to physically relevant non-linear effects such as non-linear gravity, redshift-space distortions and scale-dependent bias  \cite{2016MNRAS.455.2474A}. Crucially, these properties allow us to estimate the LP in a model-independent way from clustering data, i.e.~without resorting to a phenomenological model of the CF. Its convenient definition allows us to simply estimate the LP from the data without deriving it as a cosmology-dependent secondary parameter \cite{2018PhRvL.121b1302A}. The LP was validated with survey mocks \cite{2018PhRvD..98b3527A}, employed to estimate distances (in units of the LP) from data \cite{2018PhRvL.121b1302A}, and proven to be more constraining than the CF-MF-inferred distances (in units of the SH) \cite{2019PhRvD..99l3515A}. These convenient properties come at the cost of possible non-detection of the LP for clustering data with a low signal-to noise \cite{2018PhRvD..98b3527A} (i.e. a model-independent fit to the BAO does not force the presence of a peak and a dip in the data). This problem might be overcome  with a different LP detection definition, however the results so far obtained present this limitation. Of course, the estimate of the LP accuracy is limited by the accuracy of the N-body simulations. \\

In this manuscript we first unambiguously give the definitions of the two rulers. We then aim to completely characterize their cosmological-parameter dependence in as wide as possible a range. This is a key ingredient to assess the  constraining power of the LP and SH. This investigation is performed by means of an accurate Boltzmann code. We  then present the posteriors derived for the LP and SH, given the cosmological-parameter posteriors inferred from the CMB temperature and polarization-anisotropy power spectra as measured by the Planck satellite \cite{2016A&A...594A..13P}. \\
We shall explain how for the time being both rulers, when inferred from non-BAO observations like the CMB, need to be derived as cosmology-dependent secondary parameters. Since one of the most important uses of BAO measurements is for cosmological-model selection, this is not a limitation. In fact, to combine the PG-BAO measurements with other cosmological probes the likelihoods need to be properly combined. If we could obtain  cosmological-model-independent estimates of the rulers from non-BAO observables, it would be useful for performing cross-checks among different datasets or for deriving less model-dependent, if weaker, constraints on cosmological models.\\
Even if the CMB-derived SH and LP lengths and errors cannot be directly employed to combine CMB and BAO measurements, they are informative of the two rulers' power to  constrain cosmology and of their cosmological-model dependence. As a working example, we choose three popular cosmological models fit to the CMB data: flat-\LCDM, \LCDM$\,$ and  flat-$w$CDM. We show that, given these models, the two CMB-derived rulers have the same constraining power and are strongly correlated. This is as expected, because within $10\sigma$ from the Planck best-fit parameter values, the two rulers have the same dependence on those cosmological parameters. For larger deviations from the current CMB constraints, the two rulers keep their standard-ruler properties but present slightly different dependences on the cosmological parameters. 
Unfortunately, in the absence of N-body simulations in those regimes of cosmological parameter space, we do not know the non-linear behavior of the CF. Thus, we cannot  rely on the accuracy of these conclusions for either standard ruler. This is, however, a limitation that applies to all BAO studies unless informative priors from other cosmological probes are employed (e.g.~\cite{2015PhRvD..92l3516A}).

Interestingly, the two rulers' lengths as derived from the CMB have the same errors for all three of the cosmological models considered. This was not obvious {\it ab initio} since flat-\LCDM\, has one fewer parameter than the other two models.
We also uncover a $0.5\sigma$ shift in the best-fit length of the standard rulers in $\Lambda$CDM depending on whether one does or does not fix the curvature to be zero. This suggests that even though the lengths of the rulers are determined by early-universe physics, the estimated values of those lengths, given CMB observables, can shift between cosmological models that differ only in their late-time physics. This is because the estimated values of ``early-universe'' parameters can be affected by the differing late-universe physics.

The layout of the manuscript is as follows. In Section \ref{sec:method} we explain what we mean by a cosmological standard ruler in the context of the BAO measurements. We then provide operational definitions of the linear-point and sound-horizon rulers, and explain how their parameter dependence is investigated. In Section \ref{sec:results} we present and discuss our results, and in Section \ref{sec:conclusions} we conclude.

\section{Methodology}\label{sec:method}

\subsection{Two cosmological standard rulers: $s_{LP}$ and $r_{d}$}\label{sec:method:def}

A comoving cosmological standard ruler ${L_{{\rm sr}}}$ is a length-scale that is redshift-independent in comoving coordinates. The Purely Geometric-BAO methods rely on such a ruler. In particular, as shown in \cite{2019PhRvD..99l3515A}, from the large-scale-structure clustering data we estimate with the PG-BAO $\frac{L_{{\rm sr}}}{D_{V}(\bar{z})}$. Here $D_{V}(\bar{z})$ is the isotropic-volume distance, defined as
\be
	D_{V}(\bar{z}) \equiv \left[(1+\bar{z})^{2}D_{A}(\bar{z})^{2}\frac{c\bar{z}}{H(\bar{z})}\right]^{{1/3}}\,,
	\label{xi:y}
\ee
where $D_{A}(z)$ is the angular-diameter distance, $H(z)$ is the Hubble rate, and $\bar{z}$ is the effective redshift of the observation. To be a comoving cosmological standard ruler, we further insist that $L_{{\rm sr}}$ not depend on the primordial-perturbation parameters. 

For Purely Geometric-BAO measurements, we require, in addition, that we be able to estimate $\frac{L_{{\rm sr}}}{D_{V}(\bar{z})}$ without assuming spatial flatness or adopting any specific model for the late-time acceleration of the Universe.  In \cite{2019PhRvD..99l3515A}, two procedures and two relative rulers to estimate $\frac{L_{{\rm sr}}}{D_{V}(\bar{z})}$ from the observed CF-monopole\footnote{We recall that we work in the observed redshift space (denoted by s) where the CF is usually expanded in spherical harmonics, we consider the first term of the expansion, i.e.~the CF-monopole.} were proposed. These two rulers meet the criteria above, at least for a broad class of models that share the same functional form of the CF\cite{Anselmi:2014nya}.
The first of these procedures exploits the Linear Point standard ruler $s_{LP}(\omega_{b}, \omega_{c})$, which is defined by a specific feature of the CF computed in linear theory over the BAO range of scales. It depends only on the physical baryon and dark-matter energy densities $\omega_{b}$ and $\omega_{c}$. This feature is independent of redshift and of the primordial-fluctuation parameters. The second employs a CF-model-dependent fit that marginalizes over the primordial-fluctuation, redshift and tracer-dependent parameters. It is called CF-model-fitting (CF-MF) and relies on the comoving sound horizon at the baryon-drag epoch $r_{d} (\omega_{b}, \omega_{c})$. $s_{LP}$ and $r_{d}$ are cosmological standard rulers that we can use to estimate $\frac{L_{{\rm sr}}}{D_{V}(\bar{z})}$.

When employed to perform PG-BAO measurements, the two rulers present a key difference, rooted in the way they are estimated from data. The Linear Point can be inferred in a cosmological-model-independent way from the clustering-data, so long as a BAO dip and peak are detected in the correlation function.  The sound-horizon scale, on the other hand, is indirectly derived by mapping the cosmological parameters estimated from the correlation function (using the CF-MF for example) 
 to the SH scale through the accurate knowledge of the parameter dependence of the SH. It does not rely on the BAO dip and peak detection. To exploit cosmological-model independence of the LP, we must verify that, even though it is defined in terms of the redshift-zero CF,  the LP is actually redshift-independent, both at the linear and non-linear levels. In contrast, the sound horizon, being defined as a specific function of the cosmological parameters, is intrinsically redshift-independent.

To use the PG-BAO measurements to constrain a specific cosmological model and to break parameter degeneracies inherent in the PG-BAO outcomes, it is  standard to combine them with other observational probes. This requires that a cosmological model be adopted, and the $D_{V}(\bar{z})$ and $L_{{\rm sr}}$ dependencies on the cosmological-background parameters be specified. We thereby obtain $\frac{L_{{\rm sr}}(\omega_{b}, \omega_{c})}{D_{V}(\bar{z};\; \omega_{b}, \omega_{c}, ...)}$. 

Notice, that, while the accurate knowledge of the parameter dependence of the ruler is necessary to estimate $r_{d}/D_{V}$ from the PG-BAO, it is not needed for $s_{LP}/D_{V}$.
However, since both the ruler $L_{{\rm sr}}$ and $D_{V}$ depend on $\{\omega_{b}, \omega_{c}\}$, we need to properly combine the PG-BAO likelihood with the likelihoods from the other probes. Therefore, 
for inferring parameters, this additional feature of $s_{LP}$ is irrelevant. It could however be valuable when performing consistency tests among datasets. 

While the LP and the SH are useful for performing the PG-BAO measurements, their lengths can also be estimated from other cosmological probes (e.g.~\cite{2017JCAP...04..023V}). This allows us to perform consistency checks among different cosmological observations (e.g.~\cite{2019arXiv190307603R, 2019arXiv190710625V, 2019arXiv190711594S, 2018arXiv180706209P}). Notice that, in this case, there is no general reason why the rulers must be  independent of the cosmological model. In this sense we underline that neither the LP nor the SH can be directly estimated (i.e. without interposing a cosmological model) from non-BAO observations. Moreover this implies that there is no reason related to non-BAO observations to prefer the SH to the LP standard ruler.

In order to exploit the two rulers to learn about cosmology it is mandatory to characterize and compare them. We will first study their cosmological-parameter dependence by means of the CAMB Boltzmann code \cite{2000ApJ...538..473L, Thepsuriya:2014zda} \footnote{\url{https://camb.info}}. We then exploit this result and the CMB measurements performed by the Planck satellite to map the Planck posteriors for different cosmological models to the LP and SH posteriors. This will allow us to compare the two rulers and their errors.

\subsection{Linear Point and Sound Horizon definitions and parameter dependence}\label{sec:LPSHdef}

We start by defining the Linear Point and Sound Horizon rulers. 

The LP is defined as the mid-point between the peak and the dip of the correlation function computed in linear theory at redshift $z=0$. From a Boltzmann code (e.g.~the CAMB code), we obtain the linear matter power spectrum $P_{\rm lin}(k, z)$ at redshift $z$.  
The spatial derivative of the real space CF
\bea \label{xi:deriv}
	\xi^{{\prime}}(s, z) = -\frac{1}{2\pi^2} \int \di k \,\,k^{3}P_{\rm lin}(k, z) j_{1}(ks) \,,
\eea
where $j_{1}(x)= (- x \cos(x) + \sin(x))/x^{2}$ is the first-order spherical Bessel function. 
(For notational convenience, we denote both the real and redshift-space distance as $s$. The LP length is insensitive to which space we are considering.)
For appropriate values of the cosmological parameters, the linear correlation function (evaluated at redshift $z=0$) has a maximum (peak) and minimum (dip) in the BAO range of scales, so $\xi^{{\prime}}(s,0)$ has two zeros.
With a root-finding procedure we calculate the dip $s_{d}$ and peak $s_{p}$ positions \footnote{To ensure numerical convergence a Gaussian kernel is usually employed to perform the integral in Eq.~(\ref{xi:deriv}). To speed up the computation we use an exponential kernel $e^{-(k/k_{\rm max})^{4}}$ that gives equivalent results for a large enough $k_{\rm max}$ value. We tested that from $k_{\rm max}=1$ Mpc/h to $k_{\rm max}=100$ Mpc/h our numerical findings are unchanged. Hence we conservatively choose $k_{\rm max}=10$ Mpc/h.}, and 
\bea \label{def:LP}
	s_{LP}=\frac{s_{d}+s_{p}}{2} \,.
\eea

The comoving sound horizon is defined via \cite{Thepsuriya:2014zda}
\bea \label{def:LP}
	r_{s}(z)=\int_{0}^{\eta(z)} \frac{\di \eta^{\prime}}{\sqrt{3(1+R)}}
\eea
where $R\equiv 3 \rho_{b}/(4 \rho_{\gamma})$ is proportional to the ratio of the baryon density $\rho_{b}$ and the radiation density $\rho_{\gamma}$, and $\eta$ is the conformal time. The sound horizon at the drag epoch $r_{d}\equiv r_{s}(z_{\rm drag})$  depends on $z_{\rm drag}$.
By definition, after the drag epoch, the baryon velocity  decouples from the photon fluid, so baryon perturbations stop undergoing acoustic oscillations. 
In reality the process of decoupling is gradual, so to define $z_{\rm drag}$ one uses an indicative central value of the baryon scattering probability.  This is generally taken to be  
 $\tau_{d}(z_{\rm drag}) = 1$ \cite{1996ApJ...471..542H}, with 
\bea \label{def:tau}
	\tau_{d}(\eta)=\int_{\eta_{0}}^{\eta} \di \eta^{\prime}(\partial_{\eta^\prime}\tau_{\rm Th})/R\, \,.
\eea
where $\eta_{0}$ is the conformal time today and $\tau_{\rm Th}$ is the Thomson optical depth from recombination. 
$r_{d}$ is among the CAMB outputs. 
(Of course, the apparent dependence of $\tau_{d}(z_{\rm drag})$ on late-time baryon-photon scattering arising from the lower limit on the integral in Eq.~(\ref{def:tau}) is assumed to be subdominant.) 

As mentioned in Section \ref{sec:method:def}, we will numerically verify that the LP, estimated from the CF computed in linear theory, is redshift-independent.
This implies that both $s_{LP}$ and $r_{d}$ are given by algorithms that take as inputs only the values of the cosmological parameters. We will therefore compare the cosmological-parameter dependence of the LP and SH, demonstrating that both rulers are geometrical, i.e.~independent of the primordial-perturbation parameters, and with similar dependence on the parameters describing the background cosmology. 

To perform model-independent PG-BAO measurements, the LP also needs to be redshift independent: $s_{LP}$ extracted from $\xi^{\prime}(s,z)$ must coincide with $\xi^{\prime}(s,0)$.

\begin{figure*}
\centering
\includegraphics[width=1\hsize]{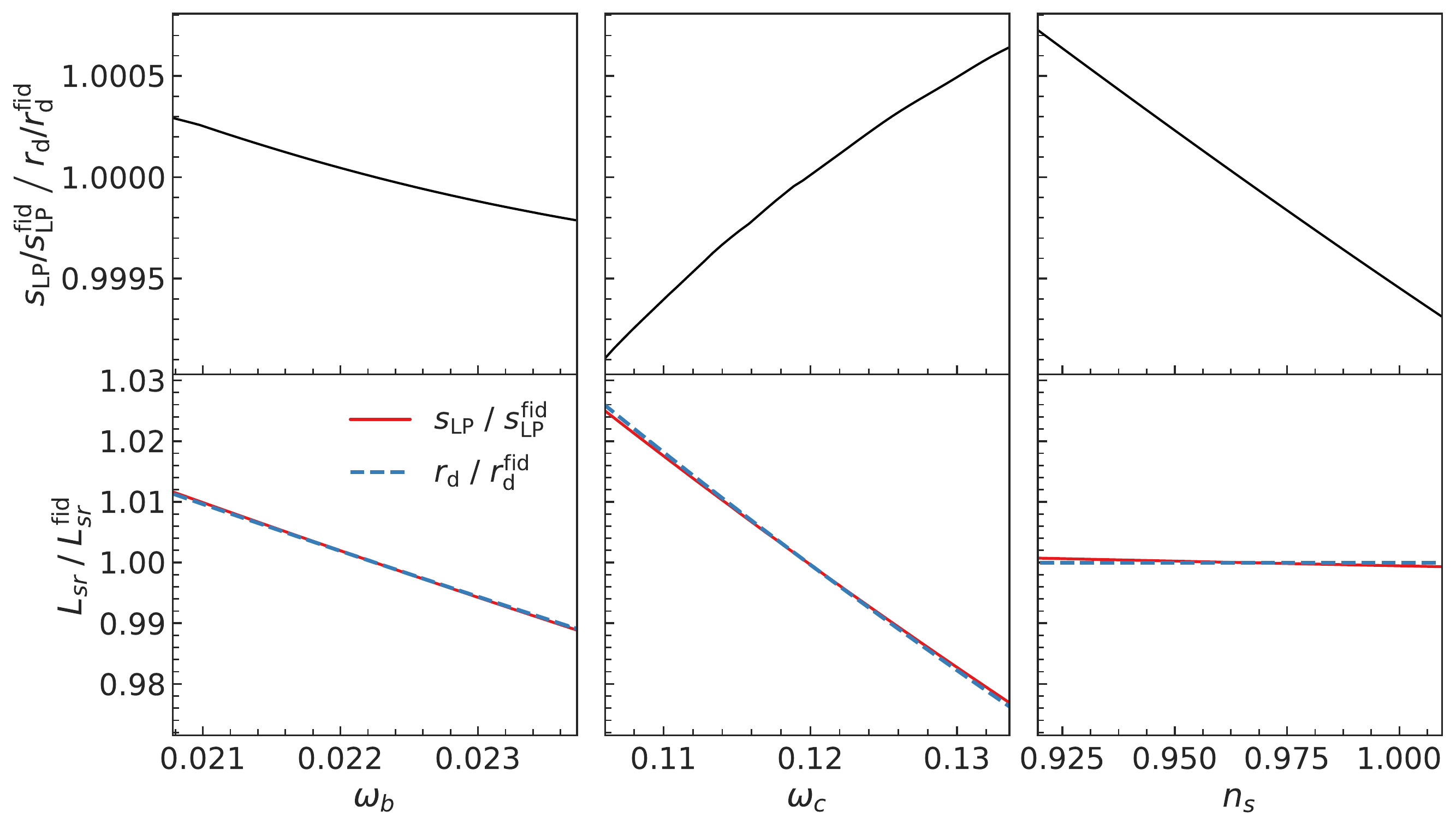}
\caption{
\label{fig:RulersAbCMB} 
{\it Linear point} and {\it sound horizon} dependence on $\omega_{b}$, $\omega_{c}$ and $n_{s}$ for a parameter range within $\pm10 \sigma$ of the Planck best-fit values for flat-$\Lambda$CDM. The rulers' lengths (indicated by $L_{\rm sr}$) are normalized, i.e.~they are each divided by their length evaluated at a fiducial value of the parameters. We plot the ratio of the normalized $s_{LP}$ to the normalized $r_{d}$ in the {\it top panels} to highlight their relative parameter dependence. In the {\it bottom panels} we present the normalized rulers as a function of the cosmological parameters.
}
\end{figure*}

\begin{figure*}
\centering
\includegraphics[width=1\hsize]{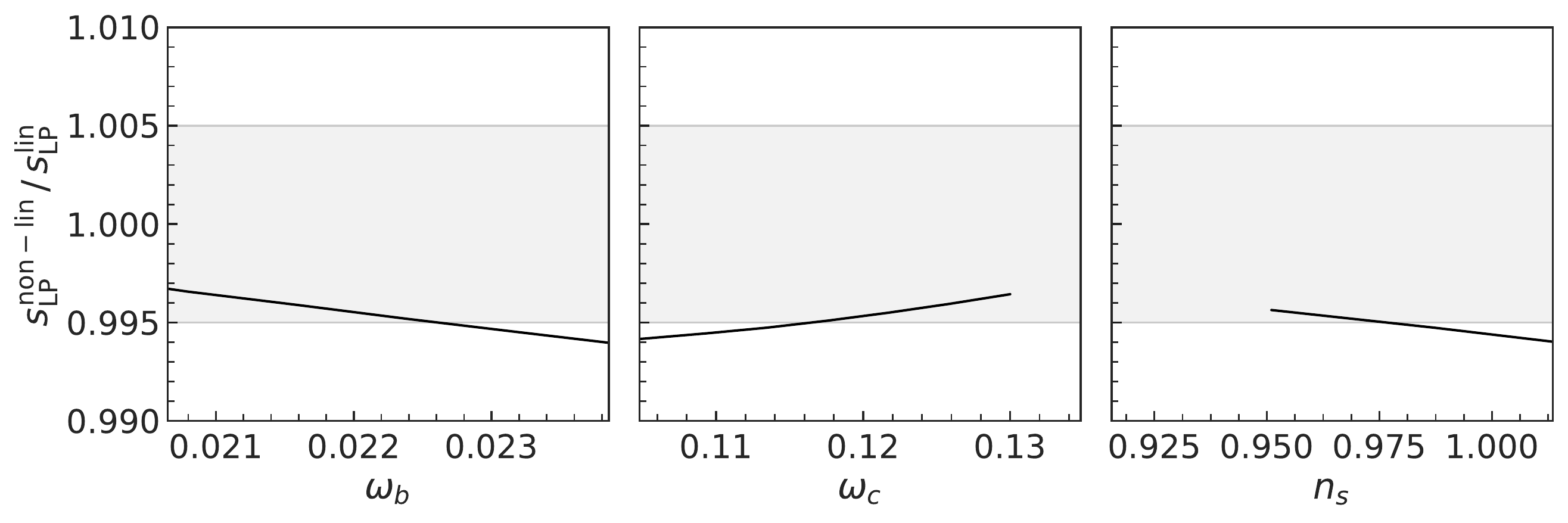}
\caption{
\label{fig:nonlinLP} 
We show the {\it linear point} inferred from the non-linear correlation function at redshift $z=0$ divided by its linear value. It is plotted in the same parameter range as Fig.~\ref{fig:RulersAbCMB} for $\omega_{b}$, $\omega_{c}$ and $n_{s}$. As expected \cite{2016MNRAS.455.2474A}, non-linear physics reduces $s_{LP}$ at z=0 to $\sim 99\%$ of its high-z value for the fiducial cosmology, motivating a $0.5\%$ correction to keep $s_{\rm LP}^{\rm non-lin}$ within $0.5\%$ of $s_{LP}^{\rm lin}$ over the full range of redshift values.
}
\end{figure*}

\subsection{Standard rulers from CMB data} 
One of the most important and constraining cosmological probes is the CMB temperature anisotropy. 
Combining the PG-BAO with CMB constraints is current standard practice (e.g.~Section 5 of \cite{2016A&A...594A..13P})\footnote{Strictly speaking, so far the BAO-only and not the PG-BAO has been employed.}. 
As mentioned above, this should be done by combining the CMB and BAO likelihoods, properly taking into account their cross correlations\footnote{While the PG-BAO is independent of primordial physics, CMB anisotropies are affected by late-time physics, e.g.~the late-time integrated Sachs Wolfe effect, Sunayev-Zeldovich effect, and CMB lensing. Moreover, both rulers enter divided by $D_V(z)$, with its parameter dependences, as discussed above.}. 
Thus estimating the best-fit value and errors of the rulers from the CMB cosmological parameter posteriors is not directly applicable for using CMB and PG-BAO jointly for cosmological parameter estimation or model selection, but it is nevertheless a valuable tool for comparing the utility of the two rulers and to uncover the cosmological-model-dependencies given the CMB constraints.

\begin{figure*}
\centering
\includegraphics[width=1\hsize]{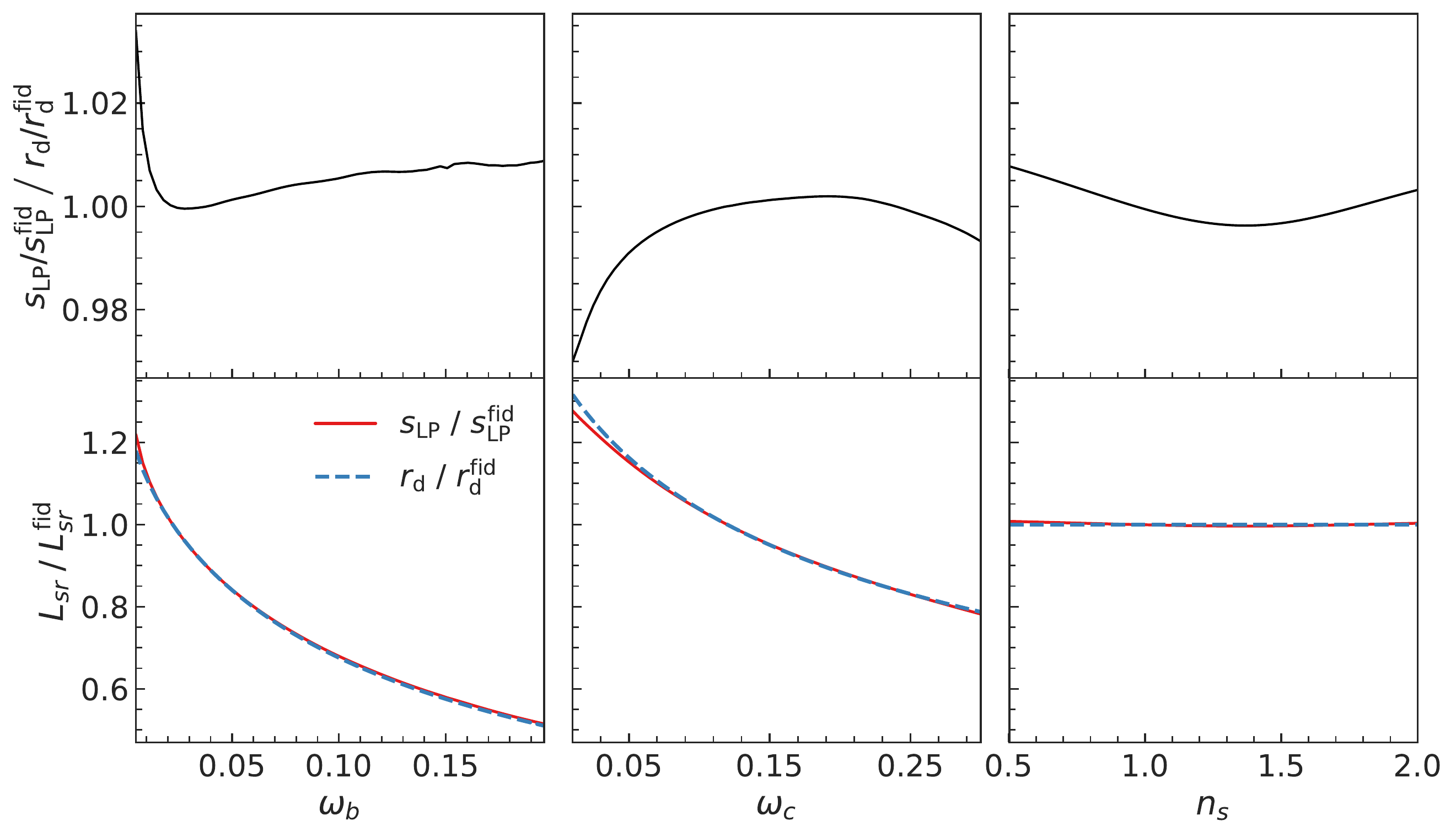}
\caption{
\label{fig:rrfid_wide} 
{\it Linear point} and {\it sound horizon} dependence on $\omega_{b}$, $\omega_{c}$ and $n_{s}$ for a wide range of parameter values. We show the ratio of the normalized $s_{LP}$ to the normalized $r_{d}$ in the {\it top panels} to highlight their relative parameter dependence. In the {\it bottom panels} we plot the normalized rulers as a function of the cosmological parameters.
}
\end{figure*}

We employ the Monte Carlo Markov chains (MCMC) released in 2015 by the Planck collaboration\footnote{\url{https://wiki.cosmos.esa.int/planckpla2015/index.php/Cosmological_Parameters}} as a result of their analysis of the Planck CMB data \cite{2016A&A...594A..13P}. We consider the chains named ``TTTEEE + lowTEB,'' in which both the low-$l$ and high-$l$ CMB anisotropy temperature and polarization power spectrum data are taken into account.

We test the standard-ruler lengths and errors derived from the CMB constraints by assuming three different cosmological models that are consistent with the PG-BAO cosmic distance measurements.\footnote{As discussed above, both PG-BAO techniques rely on certain properties of the fiducial cosmological model that are not necessarily preserved in certain alternatives, such as modified gravity theories.} Among the available Planck chains we choose flat-$\Lambda$CDM and the two simplest extensions: $\Lambda$CDM with no flatness assumption, and flat-$w$CDM (where the dark-energy equation-of-state parameter $w(z)$ is assumed to be constant).

\begin{figure}
\centering
\includegraphics[width=1\hsize]{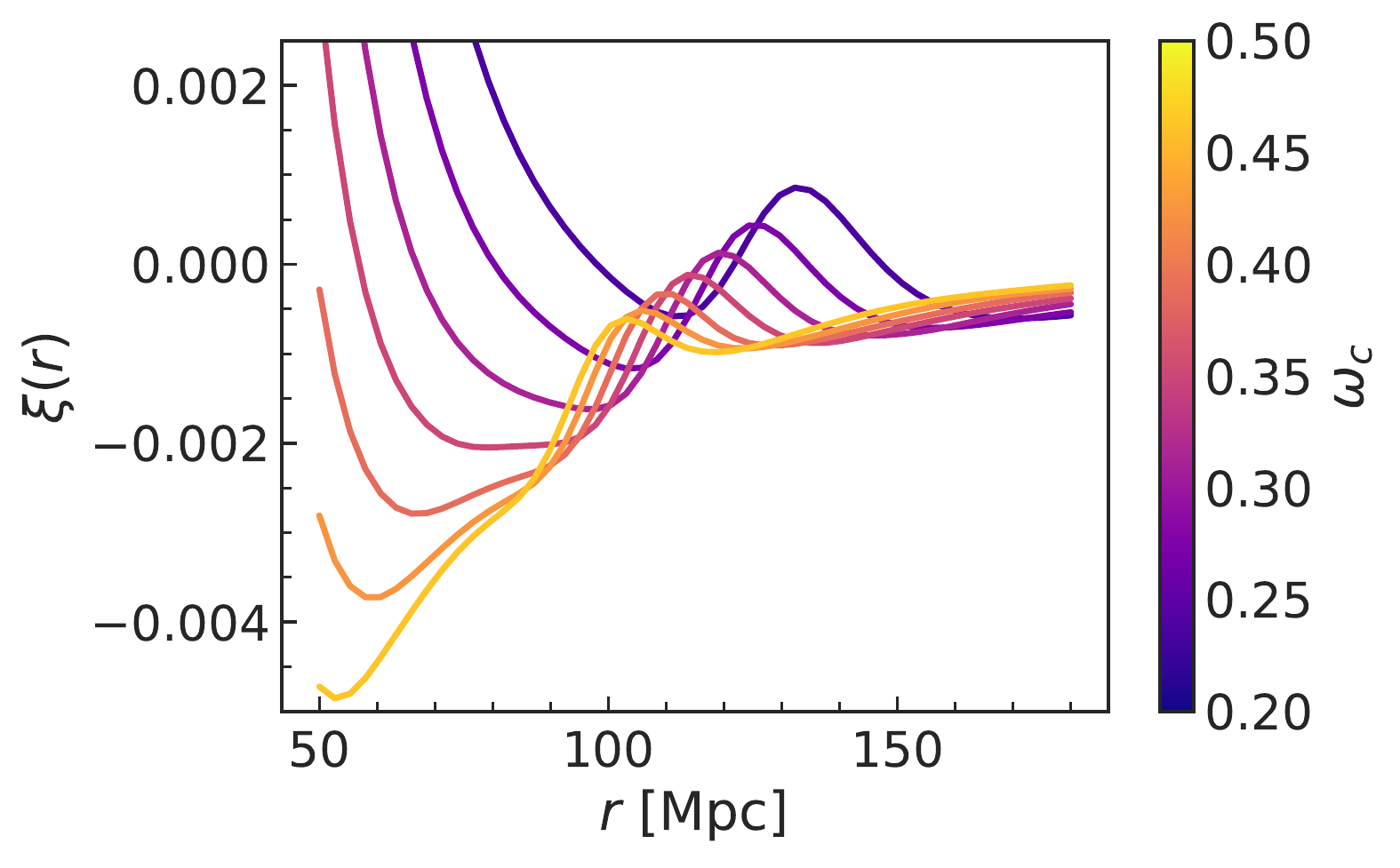}
\caption{\label{fig:xi_r_varying_omch2} 
Matter correlation function in real space (notice the use of the $r$ coordinates instead of the redshift-space  $s$  used in the rest of the paper) computed in the linear approximation. The plot shows how the CF changes with $\omega_{c}$. Notice that for $\omega_c\gtrsim0.3$ the familiar peak-dip structure of the correlation function is drastically altered. See the main text for how this behavior can be exploited to perform cosmological inference with the {\it linear point} standard ruler.
}
\end{figure}

\section{Results}\label{sec:results}
In this section we first present the results on the parameter-dependence of the standard rulers. We then compare the linear point and the sound horizon given the Planck 2015 CMB temperature and polarization power-spectrum data and a cosmological model assumed to fit them.  

\subsection{Linear Point and Sound Horizon: two geometrical standard rulers} 
We start by recalling that, to avoid introducing spurious parameter dependencies, both the linear point and the sound horizon comoving lengths need to be expressed in Mpc units and not in Mpc/h, as emphasized in \cite{2019PhRvD..99l3515A}. In these units and in the context of the PG-BAO, by definition (see Section \ref{sec:LPSHdef}) the sound horizon depends only on the physical energy densities $\omega_{c}$ and $\omega_{b}$. The LP is defined through the linear CF, and is insensitive to the optical depth parameter $\tau$, which characterizes the late-time reionization history; moreover the LP depends only on the locations of the extrema of the CF  and hence is independent of the scalar amplitude $A_{s}$ and of the Hubble constant $h$. 
Among the parameters of the standard cosmological model, it can depend only on $\omega_{c}$, $\omega_{b}$ and on the scalar spectral index $n_{s}$\footnote{The dependence of BAO-scale perturbations on curvature is negligible for any reasonable values of the curvature, ${\cal O}(0.1\%)$ for $\Omega_k<0.5$. 
With the choice of $\omega_i$ as cosmological paramters, the LP is independent of $H_0$, which only rescales the CF \cite{2019PhRvD..99l3515A}.} Since we are not aware of fundamental reasons that prevent the LP to depend on $n_{s}$ we should characterize its (in)dependence numerically by means of the CAMB code\footnote{A preliminary analysis on the $n_{s}$ linear point dependence was presented in \cite{2016MNRAS.455.2474A}.}. Finally, to compare the constraining power of the two rulers, for each of them we consider the normalized quantity $L_{{\rm sr}}/L_{{\rm sr}}^{\rm fid}$ and its parameter dependence, where the subscript ``fid'' refers to a fiducial value for the parameters that we choose to be the Planck 2015 flat-$\Lambda$CDM best fit one: $\omega_{b}=0.022252$, $\omega_{c}=0.11987$, $\tau=0.0789$, $10^{10}A_{s}=3.0929$, $n_{s}=0.96475$ and $h=0.6725$. This allows us to compare the relative functional dependences of the LP and SH.

In Fig.~\ref{fig:RulersAbCMB} we display the normalized rulers' dependence on the relevant parameters. In the top panels, to simplify their comparison, we plot the ratio of the normalized $s_{LP}$ to the normalized $r_{d}$. The rulers are plotted over a parameter range defined to be within $10 \sigma$ of the Planck best-fit values for flat-$\Lambda$CDM. This is clearly sufficient for any comparison of BAO with the CMB as measured by Planck.  In fact we are characterizing the rulers well within the CMB parameters' likelihood for all the models here considered. From Fig.~\ref{fig:RulersAbCMB} we confirm that the sound horizon is exactly geometrical, and that the linear point is geometrical at the $0.1\%$ level. Moreover, the parameter dependence of the normalized LP and SH is equivalent at the $0.1\%$ level. It is therefore reasonable to expect that the two CMB derived rulers will be strongly correlated and that their relative errors will be the same. 

In \cite{2016MNRAS.455.2474A} it was proved that the LP has a $0.5\%$ intrinsic uncertainty given by non-linear effects as non-linear gravitational evolution, redshift-space distortions and scale-dependent bias. The result was obtained employing N-body simulations, and cross-checked with a simple non-linear CF model. Since in \cite{2016MNRAS.455.2474A} the analysis was performed for one set of cosmological parameters, it is important to test here whether the $0.5\%$ intrinsic uncertainty holds within the considered $10 \sigma$ range around Planck values. To this end, we will use the simple non-linear CF model employed in \cite{2016MNRAS.455.2474A}, since we expect it to work in the parameter range considered (i.e.~close to the parameter range where the non-linear CF model has been tested). For completeness, we recall here the CF-model used in \cite{2016MNRAS.455.2474A, 2019PhRvD..99l3515A}. 

In the observed redshift space, at BAO scales, the non-linear correlation function monopole can be approximated as \cite{2019PhRvD..99l3515A}
\bea
	\xi^{\rm non-lin}_{0}(s,z) \simeq \int \frac{\di k}{k} \frac{k^{3}P_{\rm lin}(k,z)}{2 \pi^{2}} A^{2} e^{- k^{2} \sigma_{0}^{2}}\, j_{0}(ks)\,, 
	\label{nl:xi}
\eea
where we have defined
\begin{eqnarray}\label{xi:map}
	A^{2}(z)&\equiv&b_{10}^{2}+\frac{2 b_{10} f}{3}+\frac{f^{2}}{5}, \\
	\sigma_{0}^{2}(z)&=& \frac{\sigma_{v}^2 \left[35 b_{10}^2 \left(f^2+2 f+3\right)+14 b_{10} f \left(3 f^2+6 f+5\right)\right.}{105 \,A^{2}}  \nn \\
	&+&\frac{\left.3 f^2 \left(5 f^2+10 f+7\right)\right]}{105 \,A^{2}} 
		-\frac{2 b_{01}(3 b_{10}+f)}{3\,A^{2}}\,\nn \,.
\end{eqnarray} 
Here
$P_{\rm lin}(k,z)$ is the linear matter power spectrum (PS) at redshift $z$;
$b_{10}$ is the Eulerian linear bias and $b_{01}$ is the scale-dependent bias;
$f(z)=\di \ln D/\di \ln a$ is the growth rate at redshift $z$;
$j_{0}(x)=\sin(x)/x$ is the zero-order spherical Bessel function; and 
$\sigma_{v}(z)$ is the one-dimensional dark-matter velocity dispersion in linear theory, given by\be
	\sigma_{v}^{2}(z)=\frac{1}{3} \int \frac{\di^{3} q}{(2 \pi)^{3}}\frac{ P_{\rm lin}(q,z)}{q^{2}} \, .
	\label{sigma:v}
\ee  

In Fig.~\ref{fig:nonlinLP} we plot the linear point inferred from the non-linear correlation function at $z=0$ divided by its linear value, over the Planck $10\sigma$ range.
$s^{{\rm non-lin}}_{LP}$ is approximated 
	by equation (\ref{nl:xi}), 
augmented by our standard $0.5\%$  non-linear correction \cite{2016MNRAS.455.2474A}. 
We see that the linear point basically remains within $0.5\%$ 
of its linear value wherever it is defined.
Where the peak and dip are absent, $s_{LP}$ is undefined.
We expect some similar (if less pronounced) difficulty extracting $r_d$ from CF-MF.
Notice that redshift zero represents the worst case, 
with the most non-linear evolution, and that all real cosmological measurements are performed at higher redshift
where the problem is less severe.

The PG-BAO are meant to be used in combination with all the other cosmological observables, not only the CMB. Consequently, we should not restrict our characterization of them to within a few sigmas of the Planck results, as this could potentially bias the cosmological parameter estimation or model selection. Therefore we characterize the standard rulers over a more extended range of parameter values. This is suggested by the findings of \cite{2019PhRvD..99l3515A} where the marginalized posteriors of the cosmological parameters from the PG-BAO were found to be very wide. 

Before presenting this analysis, we warn the reader that, in the extended parameter range we will consider, there are no (or at least no accurate) N-body simulations available for us to test the non-linear effects on the correlation function in the BAO range of scales. This is an important issue for all PG-BAO analyses. In fact, the simulations used to calibrate the analysis methods and to analyze the data are performed over a much narrower range of cosmological parameters than is justified by either the prior or the posterior. Hence, the usual attitude is to assume that the BAO method works in regimes where it has not been tested. As extrapolating non-linear effects is very likely not a safe practice, the results presented in the extrapolated regions are probably not reliable. The same issue clearly applies to the Linear Point. To be specific we do not know whether the maximum $0.5\%$ shift given by non-linear effects would hold outside of the $10\sigma$ Planck parameter region. For the same reason we cannot apply the non-linear model that we employed to test non-linear effects in the $10\sigma$ Planck parameter region.

In Fig.~\ref{fig:rrfid_wide}, we display the rulers' dependence for $0.5<n_{s}<2$. The $n_{s}$ dependence of the LP is still within the $0.5\%$ LP intrinsic uncertainty \cite{2016MNRAS.455.2474A}, 
and can thus be neglected in the cosmological analysis. 
For $n_{s}>2$, this no longer the case. We take $n_{s}>0.5$ because for $n_{s}<0.5$ the BAO feature is erased. This is not a problem, as to estimate the LP from the data we employ a procedure that is independent of both any two-point-CF model and of the cosmology. We can therefore perform a conditional cosmological analysis: if the LP is detected then $n_{s}>0.5$\footnote{There is also a quantifiable probability of a false positive detection given by the effects of cosmic and sample variance.}.  This has the advantage of providing an immediate lower limit on $n_s$ no matter the {\it{ab initio}} prior.  There is no similar upper bound on $n_s$ in the range we explored. 
The CF-MF has no such opportunity to simply constrain the range of $n_s$. For this reason we explore the minimum value of $n_{s}$ allowed by the CAMB code. We found that for $n_{s}\leq0$ the code issues a warning and terminates\footnote{This holds at least for pycamb 1.0.6, the code we are employing in our analysis, i.e.~the python version of the CAMB code.}; we therefore consider only $n_{s}>0$.

We note that Boltzmann codes like CAMB will necessarily be designed and validated for specific ranges of the cosmological parameters. These may well not include values that become of interest to cosmologists, especially when there are tensions among cosmological parameters extracted from different sets of cosmological observables.  It is the responsibility of the user to rely on the code only where it is reliable.

\begin{figure}
\centering
\includegraphics[width=1\hsize]{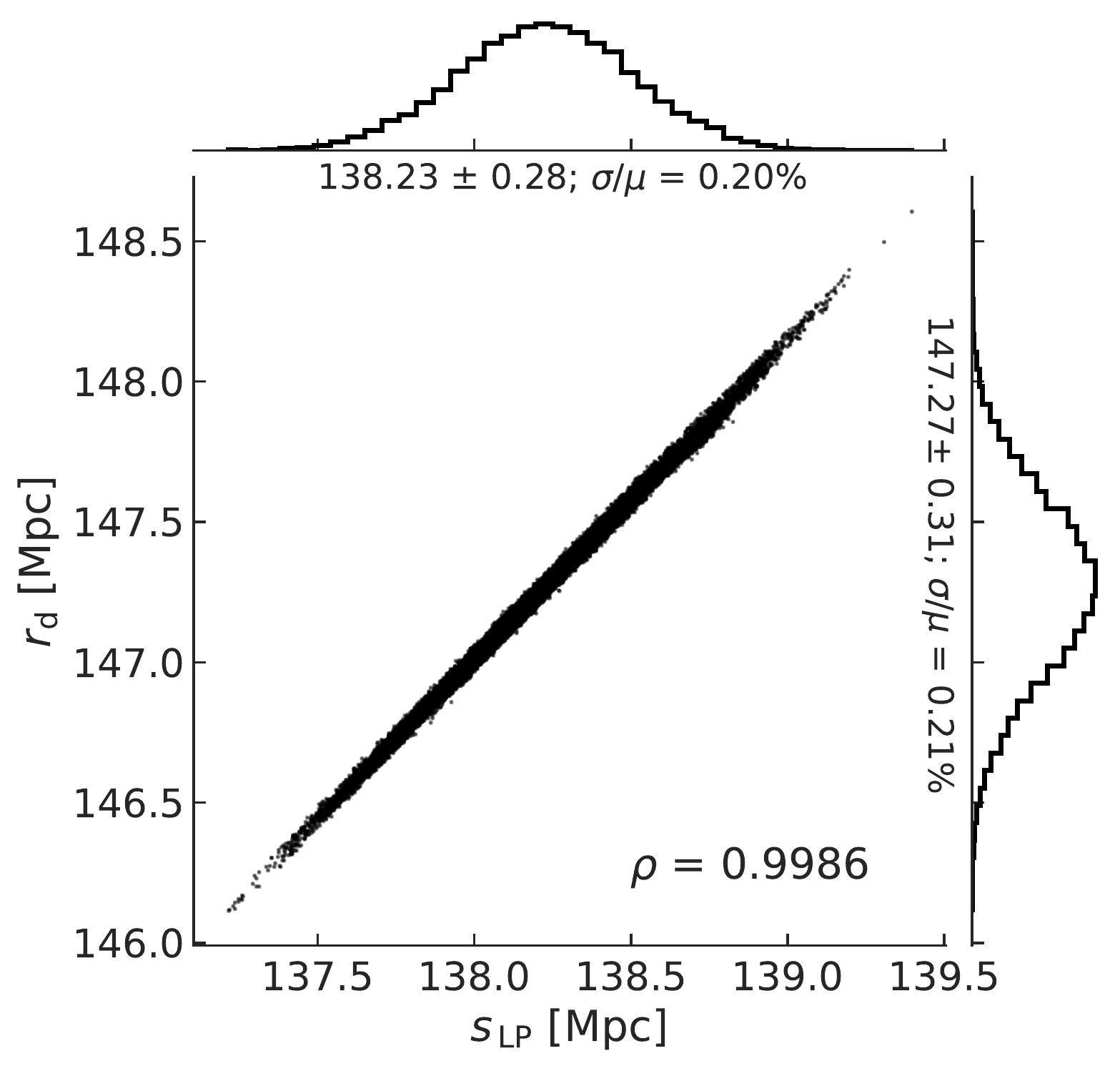}
\caption{
\label{fig:rdragvsrlpcmb} 
{\it Linear point} and {\it sound horizon} scatter plot given the Planck team's CMB cosmological constraints for the flat-\LCDM\, model \cite{2016A&A...594A..13P}. The Pearson correlation coefficient $\rho_{{\rm flat-\Lambda CDM}}=0.9986$ shows that $r_d$ and $s_{LP}$ are extremely correlated. The two rulers have the same relative errors in flat-\LCDM -- they are equivalent standard rulers from the CMB point of view.
}
\end{figure}

We should acknowledge that the viewpoint that $n_s$ is a constant for all $k$ (no running) is an approximation that cannot hold exactly and that seems more likely to fail the larger the value of $\vert n_{s}-1\vert$.  
This certainly is true within the context of slow-roll inflation, where $\vert n_{s}-1\vert$ is linear in slow-roll parameters, and the running is quadratic.
If $\vert n_{s}-1\vert>1$ then the slow roll approximation is not even valid.
Furthermore, we are measuring $n_s$ over a narrow range of length (and so momentum) scales, making global arguments about the value of $n_s$ over the full range of momenta less clearly applicable.
It is thus enormously difficult to put well-motivated priors on the space of phenomenological paramters like $n_s$ outside the very narrow range near $n_s=1$.  This presents a considerable challenge to CF-MF, if it wishes to explore
a wide prior for $n_s$; and complicates the use of both LP and CF-MF for cosmological parameter determination far outside the usual range of $n_s$ and running.

\begin{figure*}
\centering
\includegraphics[width=1\hsize]{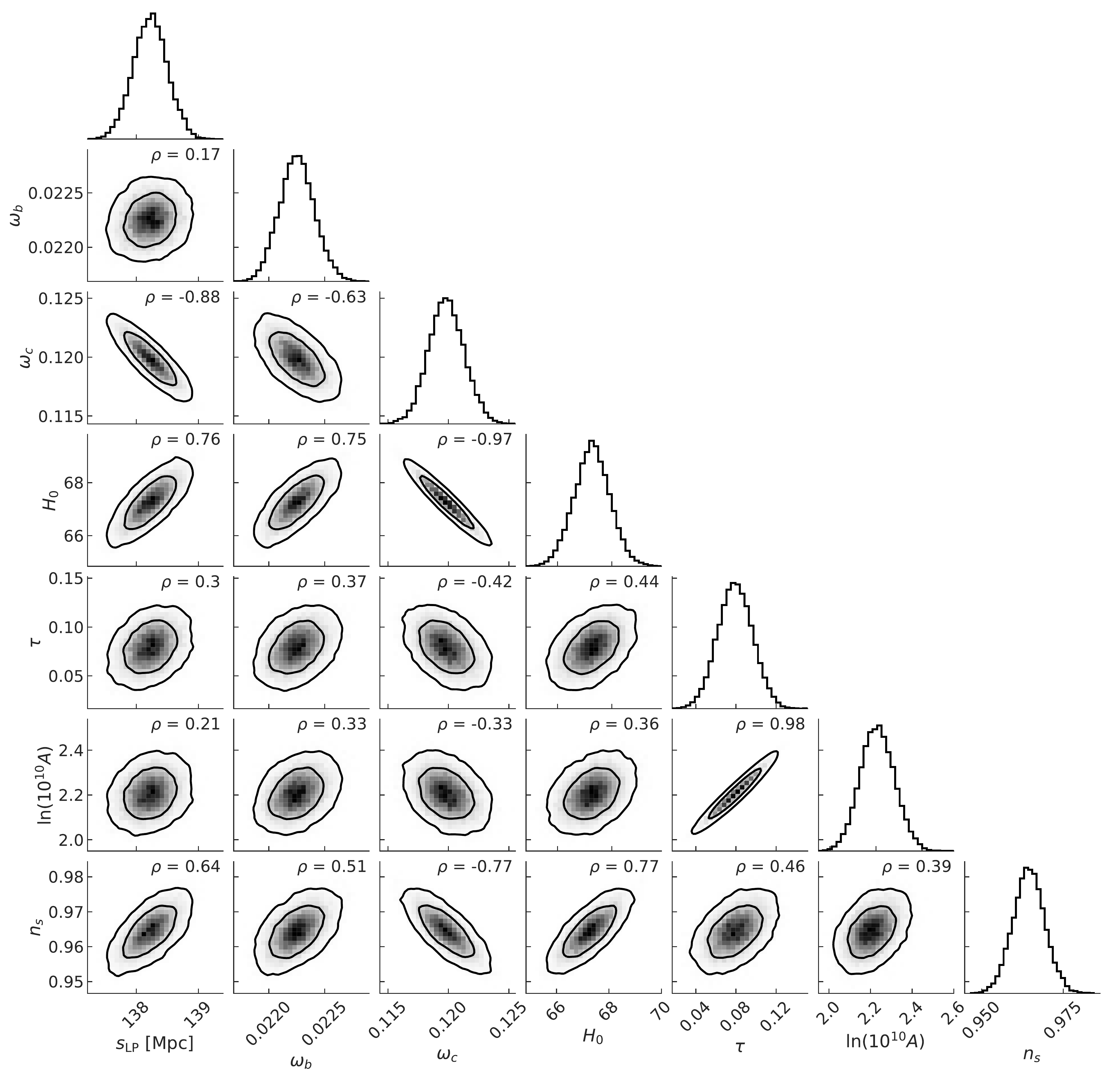}
\caption{
\label{fig:cmbcorrelationmatrix} 
Marginalized posteriors for the {\it linear point} combined with the cosmological parameters given the CMB constraints. The reported correlations are given both by the parameter dependence of the LP and by the covariance of the CMB power-spectra.  See the main text for a discussion of these correlations.
}
\end{figure*}

In Fig.~\ref{fig:rrfid_wide}, we also display the dependence of $L_{{\rm sr}}/L_{{\rm sr}}^{\rm fid}$ as a function of $\omega_b$ and $\omega_c$, for both standard rulers. We first observe that both standard rulers are usable over a wider range of these cosmological parameters than just the 10$\sigma$ Planck range used above. Moreover, they have very similar dependences over this range, i.e.~within about 3\% of each other. That they are not exactly equivalent has no particular significance for their relative utility.

The upper limit of $\omega_b<0.196$ arises from a failure of pycamb\footnote{This might be related to the CAMB version we are using (pycamb 1.0.6) as we found that an older Fortran version of the code (year 2015) works for $\omega_b<0.45$.}. This failure could represent a  challenge for the CF-MF in the absence of a prior that excludes problematic values as CF-MF can only be applied within the regime where the Boltzmann code can be trusted. The LP suffers from no such constraint -- as long as there is a peak and a dip in the data, the value of the LP can be extracted. Its use for cosmological parameter estimation or model selection, does however rely on a Boltzmann code that has been validated over the relevant region of parameter space.  

\begin{figure*}
\centering
\includegraphics[width=1\hsize]{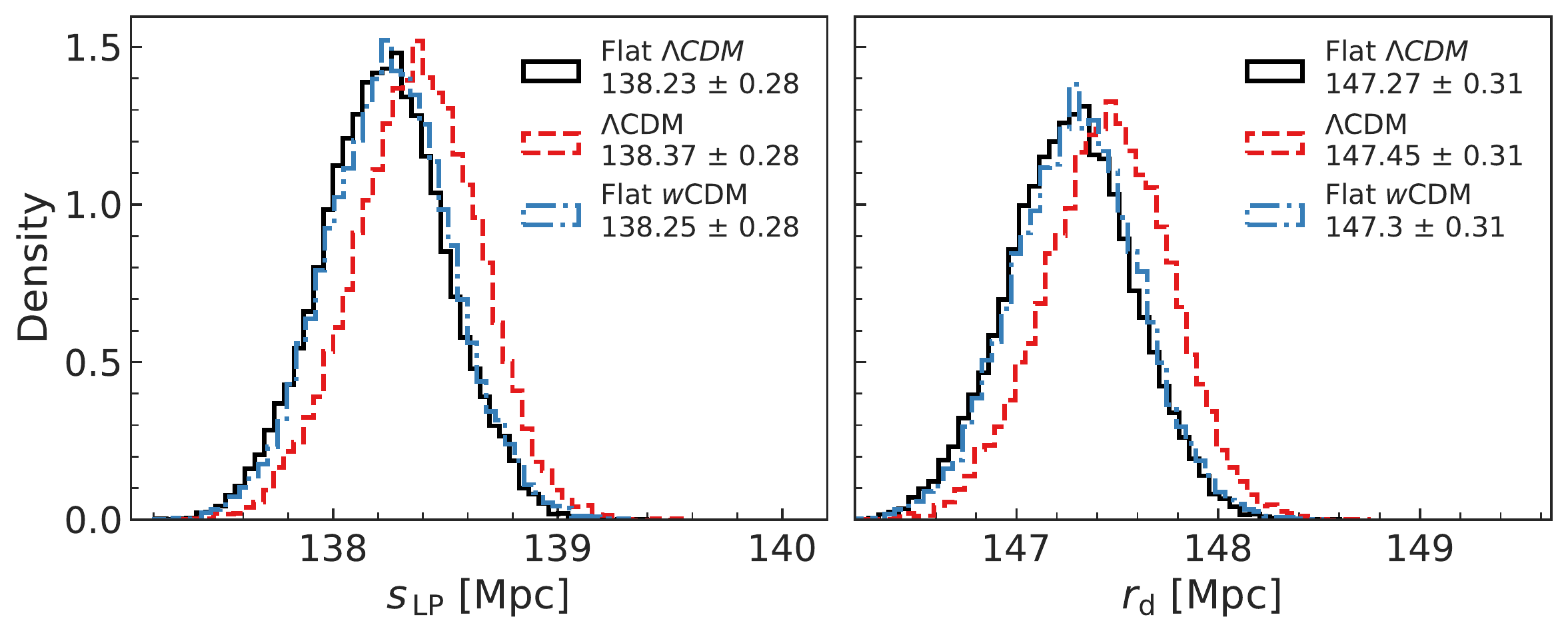}
\caption{
\label{fig:lp_dist_joint} 
{\it Linear point} and {\it sound horizon} marginalized probability distributions derived from the CMB constraints. The results are shown for three cosmological models encompassed by the PG-BAO and analyzed by the Planck team: flat-\LCDM, \LCDM$\,$ and  flat-$w$CDM.
}
\end{figure*}

For $\omega_c\gtrsim0.3$, the familiar peak-dip structure of the correlation function is drastically altered (as seen in Fig. \ref{fig:xi_r_varying_omch2}) -- the peak remains, but a new dip emerges at much smaller scales $s\lesssim70$ Mpc. We have not validated the sensitivity of the linear point to non-linear physics in this regime, and cannot do so without appropriate simulations. This suggests that the appropriate approach is, as for $n_s$, we perform an analysis conditioned on $s_{d}\gtrsim95$ Mpc, appropriate for $\omega_c\leq0.3$. If a dip is detected at  $s_{d}\lesssim95$ Mpc, then we learn that
$\omega_c\geq0.3$. Finally, since it is important for the CF-MF, we tested the maximum $\omega_{c}$ value allowed by CAMB. We found that for $\omega_{c}> 2.311$ CAMB reports a numerical issue. This is probably a good enough upper limit to set the CF-MF prior but it should be carefully tested case by case for each galaxy survey and chosen redshift bin.

There is no physical lower limit on $\omega_b$ or $\omega_c$, except of course that $\omega_b>0$ (galaxies exist) and $\omega_c\geq0$. We tested that the CAMB version we are using (pycamb 1.0.6
) works for $\omega_{b}\geq 0.00196$ and $\omega_{c}\geq 0$. Both the CF-MF analysis and the LP interpretation must deal with this probably artificial lower limit of CAMB.

A complete LP analysis would vary  all of $\{\omega_{c},\omega_{b},n_{s}\}$ simultaneously. Since the standard rulers are found to have only weak dependences on the parameters over the $10\sigma$ range around the Planck best fit values, this is not needed for comparing with CMB data, but would be necessary for a full characterization of the PG-BAO.
We leave this more complete but involved study for future work.

Finally, in order for the LP to be employed as a model-independent PG-BAO standard ruler, we need to verify numerically that, in linear theory the LP is redshift independent. We verified that this holds true in flat-\LCDM\, at the $0.1\%$ level for $z\leq10$. We did not perform the computation for higher redshift values because the lower limit of the integral appearing in Eq.~(\ref{xi:deriv}) is in principle $k=0$. This suggests a problematic dependence of $\xi^{{\prime}}(s)$ on long-wavelength gauge-dependent modes. Fortunately, for $z\leq10$, these long-wavelength modes contribute negligibly (less than 0.1\%) to the integral. However, this is indeed an issue in the BAO range of scales for larger redshifts, such as may one day be accessible to tomographic surveys.  We will address this question in a future publication.

\subsection{The lengths of standard rulers from the CMB}

As promised above, we employ the 2015 Planck ``TTTEEE + lowTEB,'' MCMC chains \cite{2016A&A...594A..13P} to determine the
the standard-ruler lengths and errors derived from the CMB constraints. We can also test their robustness by assuming three different cosmological models that are consistent with the PG-BAO cosmic distance measurements: flat-$\Lambda$CDM, $\Lambda$CDM with no flatness assumption, and flat-$w$CDM (where $w(z)$ is assumed to be constant).

In Fig. \ref{fig:rdragvsrlpcmb}, we show that $r_d$ and $s_{LP}$ are extremely correlated (Pearson correlation coefficient $\rho_{{\rm flat-\Lambda CDM}}=0.9986$) with same relative errors in flat-\LCDM. They are equivalent standard rulers from the CMB point of view. Therefore for further studies with the CMB and flat \LCDM\, we consider only $s_{LP}$ that has been so far much less investigated than $r_{d}$.

In Fig. \ref{fig:cmbcorrelationmatrix} we examine the
correlations between $s_{LP}$ and the cosmological parameters
(and the correlations among those parameters). We observe that
$s_{LP}$ is strongly anti-correlated with $\omega_c$, as expected.  The anti-correlation is less than perfect because of the dependence of $s_{LP}$ on $\omega_b$ -- the variation of $\omega_b$ for fixed $\omega_c$ somewhat washes out the $s_{LP}$-$\omega_c$ correlation. Meanwhile, $s_{LP}$ appears uncorrelated with $\omega_b$, because $\omega_c$  has a larger fractional error than $\omega_b$. Variation in $\omega_c$ for fixed $\omega_b$ washes out the underlying dependence of $s_{LP}$ on $\omega_b$ in the MCMC chain.

$s_{LP}$ is also strongly correlated with $H_0$, because $H_0$ is strongly correlated with $\omega_c$ (and $\omega_b$). 
Similarly, $s_{LP}$ is mildly correlated with $n_s$, because $n_s$ is mildly correlated with $\omega_c$ (and  $\omega_b$). 

In Fig. \ref{fig:lp_dist_joint} we examine the probability distribution function of $s_{LP}$ and $r_{d}$ for flat-\LCDM, as well as for \LCDM\, without the flatness constraint, or allowing the value of $w$ to vary away from $-1$ (but remain constant). We learn that $s_{LP}$ and $r_{d}$ errors are insensitive to variation of the curvature or of the dark-energy equation of state. This suggests that the parameters characterizing the model extensions are at most weakly correlated to the rulers given the CMB spectra and relative covariances. We tested that this is indeed the case: $\omega_{b}$ and $\omega_{c}$ are weakly correlated to $\Omega_{k}$ and uncorrelated to $w$. However, we are much more sensitive to small changes in the best-fit values of the rulers, and indeed find a shift of $0.5\sigma$ between flat-\LCDM~ and \LCDM\footnote{This is in contrast to \cite{2017JCAP...04..023V}, which appears to find that allowing for non-zero curvature causes no increase in the most likely value of $r_d$ but  introduces skewness in the PDF for $r_d$ that increases its median value. However, this difference may be a consequence of their approach, which marginalizes over late-time physics.}, 
suggesting that the LP and SH CMB-derived lengths are cosmological-model dependent already for simple extensions to \LCDM\, such as those considered here. In \cite{2019ApJ...874....4A} the authors report a slightly different result, with an even larger discordance. This might be due to a different set of Planck data used (which is not precisely reported in the manuscript). However, unlike us, they considered the cosmological-model dependent standard ruler shift irrelevant. We leave a careful investigation of this subject for the future. 
  
Finally we find that, similarly to the flat-\LCDM\, case, the LP and SH are extremely correlated : $\rho_{\rm \Lambda CDM}=0.9969$ and $\rho_{\rm flat-wCDM}=0.9984$.  The PDFs of $s_{LP}$ and $r_{d}$ are mildly non-Gaussian, but at a level that can only be detected with over $30000$ MCMC points at a p-value of $0.01$.

\section{Conclusions}\label{sec:conclusions}
BAO distance measurements are crucial to the search for which cosmological model best describes our Universe. They are a key observable for tracking the time evolution of dark energy and thereby shedding light on its nature. The Purely Geometric-BAO methods \cite{2019PhRvD..99l3515A} are a set of very attractive approaches to the BAO. They allow us to infer cosmic distances in units of a comoving cosmological standard ruler encompassing both \LCDM\, and quintessence models, without assuming a specific spatial curvature for the Universe. Furthermore PG-BAO outcomes are independent of the primordial cosmological parameters.\\
Two Purely Geometric BAO methods have been identified to date \cite{2019PhRvD..99l3515A}. The first method is CF-Model-Fitting, which relies on a CF-model template that is fit to the data. CF-MF  exploits the Sound Horizon standard ruler, which is derived as a secondary parameter from the value of the energy densities that are inferred from the fit. The second method is based on the Linear Point standard ruler \cite{2016MNRAS.455.2474A, 2018PhRvL.121b1302A, 2018PhRvD..98b3527A}. Its properties allow one to estimate the ratio of the LP to the cosmic distance from the clustering data in a model-independent way. 
The two approaches and the two relative rulers thus differ crucially in how they are extracted from galaxy data. 
  
Particularly if one aims to combine the PG-BAO measurements with other cosmological probes, it is necessary to compare the rulers' parameter dependence. This is the first analysis we present in this paper. In addition we estimate the best fits and errors of the LP and SH when they are estimated from CMB anisotropies, one of the most important cosmological observations complementary to BAO.

We find that, within $10\sigma$ of the Planck best-fit values in flat-\LCDM, at the $0.1\%$ level of accuracy, the two rulers show the same parameter dependence and  are both geometrical. This analysis is enough to interpret the CMB-derived ruler lengths. However the PG-BAO measurements need to be flexible enough to be combined with other cosmological datasets. Since a generic dataset can extend its parameter range well beyond $10\sigma$ from Planck (e.g. the Type Ia supernovae measurements \cite{2016NatSR...635596N}), we investigate the rulers' dependence for the whole allowed range of parameter values; 
We caution the reader that, in the absence of 
N-body simulations to calibrate the results in this range 
of parameter values, it is difficult to assess the reliability of any conclusions -- a problem  overlooked so far in all the cosmological analyses that employ BAO distance measurements (e.g. \cite{2015PhRvD..92l3516A})

We find that we cannot explore $\omega_{b}>0.196$ because of an internal numerical error of the CAMB Boltzmann code. For low values of the scalar spectral index, i.e.~$n_{s}<0.5$ the BAO feature is washed out. 
Therefore if we detect the LP in galaxy data we can immediately conclude that $n_{s}>0.5$. 
For $\omega_{c}>0.3$, the peak-dip structure necessary to define and detect the LP is altered and, since we did not test the LP with high-resolution N-body simulations in this parameter range, we are not allowed to interpret the results.
However, this result suggests that if the detected dip scale is smaller than $\sim 95$ Mpc, then we can conclude that $\omega_{c}>0.3$. 
In the parameter range where we could compare the two rulers, they show a similar parameter dependence, and thus have similar constraining power when estimated from non-BAO observables.
Note that we have only varied one parameter at a time, keeping  the other parameters fixed; 
a more comprehensive analysis would vary all of 
$\{\omega_{c},\omega_{b},n_{s}\}$ simultaneously, extending our findings to a three-dimensional parameter space. 
However, since the CMB Planck posterior is highly informative it has very small errors, hence this investigation is not mandatory for the present manuscript and we leave it for the future.  

We finally compute the LP and SH best fit and errors given the parameter posterior of the Planck team. 
We do this exercise for three simple cosmological models encompassed by the PG-BAO: flat-$\Lambda$CDM, $\Lambda$CDM with no flatness assumption, and flat-$w$CDM. 
We obtain that the best fits shift by $0.5\sigma$ for $\Lambda$CDM, showing that the rulers, when estimated from a particular observable as secondary parameters, could be sensitive to the assumed cosmological model, even for simple extensions such as we consider in this manuscript. Therefore, contrary to common intuition, even if the physical processes that underpin ruler are insensitive to a specific cosmological parameter (e.g.~the spatial curvature), that parameter can nevertheless influence the estimated value of the ruler due to the correlations between parameters intrinsic to specific cosmological observables. 

 We find that the LP and SH are always extremely correlated (Pearson correlation coefficient $\sim 0.999$). 
 For all three models the errors are always $0.2\%$ both for the LP and the SH. 
 
The present analysis assumes that neutrinos are massless. 
It is thus important to investigate, by means of dedicated N-body simulations and analytical approaches \cite{2015JCAP...07..043C, 2016JCAP...07..034C, 2015JCAP...07..001P}, whether or not the LP retains its PG-BAO properties when neutrinos are massive. A second important step consists in extending the analysis we presented here to the case of massive neutrinos \cite{Thepsuriya:2014zda}.

We show that the LP and SH standard rulers inferred from the CMB are equivalent for certain simple extensions of the flat-\LCDM\, paradigm. It is thus interesting to employ the two rulers estimated from different BAO observations and check their consistency in the context of the PG-BAO approaches. Moreover the behavior of the two rulers should be explored for modifications of  early Universe physics. This would allow us to further extend the consistency checks among cosmological datasets, an important investigation in this era of data-driven cosmology (e.g.~\cite{2019arXiv190307603R, 2019arXiv190710625V, 2019arXiv190711594S, 2018arXiv180706209P,2019arXiv190808950M}).

\begin{acknowledgments} 
S.A. thanks Suvodip Mukherjee for discussions on physical parameter constraints and on the meaning and usage of prior information. We thank Ariel Sanchez for discussions on parameter priors and posteriors in BAO fitting methods. We thank Anthony Lewis for feedback on the CAMB code $\omega_{b}$ upper limit. GDS was partially supported by a Department of Energy grant DE-SC0009946 to the particle astrophysics theory group at CWRU. 
The research leading to these results has received funding from the European Research Council under the European Community Seventh Framework Programme (FP7/2007-2013 Grant Agreement no. 279954) ERC-StG "EDECS". IZ is supported by National Science Foundation grant AST-1612085.
This research was supported by the Munich Institute for Astro- and Particle Physics (MIAPP) of the DFG cluster of excellence "Origin and Structure of the Universe."
This work made use of the High Performance Computing Resource in the Core Facility for Advanced Research Computing at Case Western Reserve University.
\end{acknowledgments}

\bibliography{MyBib}

\end{document}